\documentstyle[aas2pp4,11pt]{article}

\lefthead {Kurtz and Mink}
\righthead {Eigenvector Sky Subtraction}

\begin {document}
\title{Eigenvector Sky Subtraction}

\author{Michael J. Kurtz  and Douglas J. Mink}
\affil{Harvard-Smithsonian Center for Astrophysics, Cambridge, MA 02138}

{\hfil Accepted for publication in The Astrophysical Journal (Letters)
2000 March 7. \hfil}

\begin {abstract}

We develop a new method for estimating and removing the spectrum of
the sky from deep spectroscopic observations; our method does not rely
on simultaneous measurement of the sky spectrum with the object
spectrum.  The technique is based on the iterative subtraction of
continuum estimates and Eigenvector sky models derived from Singular
Value Decompositions (SVD) of sky spectra, and sky spectra
residuals. Using simulated data derived from small telescope
observations we demonstrate that the method is effective for faint
objects on large telescopes.  We discuss simple methods to combine our
new technique with the simultaneous measurement of sky to obtain sky
subtraction very near the Poisson limit.

\end {abstract}


\section {\label {intro} Introduction}

Light from the night sky interferes with the analysis of the spectra
of all but the brightest astronomical objects.  For the determination
of galaxy redshifts this limit was reached about forty years ago
(\cite{1956AJ.....61...97H}).  With the advent of digital
spectrographs (e.g. \cite{1975ApJ...197L..95W},
\cite{1976PASP...88..960S}) simultaneous measurement of object and
night sky spectra became routine, thus allowing sky subtraction.  The
basic principle of sky subtraction has been that the sky spectrum
should be observed in as identical a manner as possible (in time,
location on the sky, and optical path) to the sky spectrum observed in
the direction of the object.  As ever fainter objects have been
observed the accuracy of the sky subtraction has become more critical,
and increasingly sophisticated methods, mechanical, observational, and
algorithmic, have been employed to match the observed sky to the sky
in front of the object.  For wide field, multiobject, fiber optic
spectrographs this procedure has been especially problematic.  \cite
{1998fopa.proc...50W} review the problem of sky subtraction with fiber
optic spectrographs in detail.

In this paper we present a new method for removing the influence of
the night sky emission from the spectra of faint astronomical objects;
this new method does not rely on the simultaneous measurement of the
sky with the object.

\section {\label {iter} Iteratively Estimating the Sky}

Removing the sky from an object $+$ sky observation is equivalent to
estimating the sky spectrum in the direction of the object at the time
of observation.  Because the sky spectrum is composed of several
components the first step in the estimation process must be to decide
which of these components must be accurately estimated, and which can
be.

Dark sky spectra may be characterized by a smooth continuum, plus weak
absorption lines (from the zodiacal light) and a host of emission
lines.  For many projects involving multi-fiber spectroscopy of faint
galaxies (e.g. redshift surveys) the continuum of the object is
removed before the analysis (\cite {1998PASP..110..934K}, hereafter
KM98); in these instances it is not necessary to estimate the
continuum of the sky separately; it suffices to estimate the continuum
of the object $+$ sky.  Additionally, some lines in the sky spectrum
are so bright that errors in their estimation, which are large compared
with typical features in the object spectra, are unavoidable.

Our new method for sky subtraction begins by estimating and
subtracting the sky $+$ object continuum; this step removes typically
more than 80\%\ of the sky flux, as well as the object's continuum.
All through the sky subtraction process, we ignore pixels which
correspond to the bright sky lines we choose not to estimate.

\begin{figure}[t]
\plotone{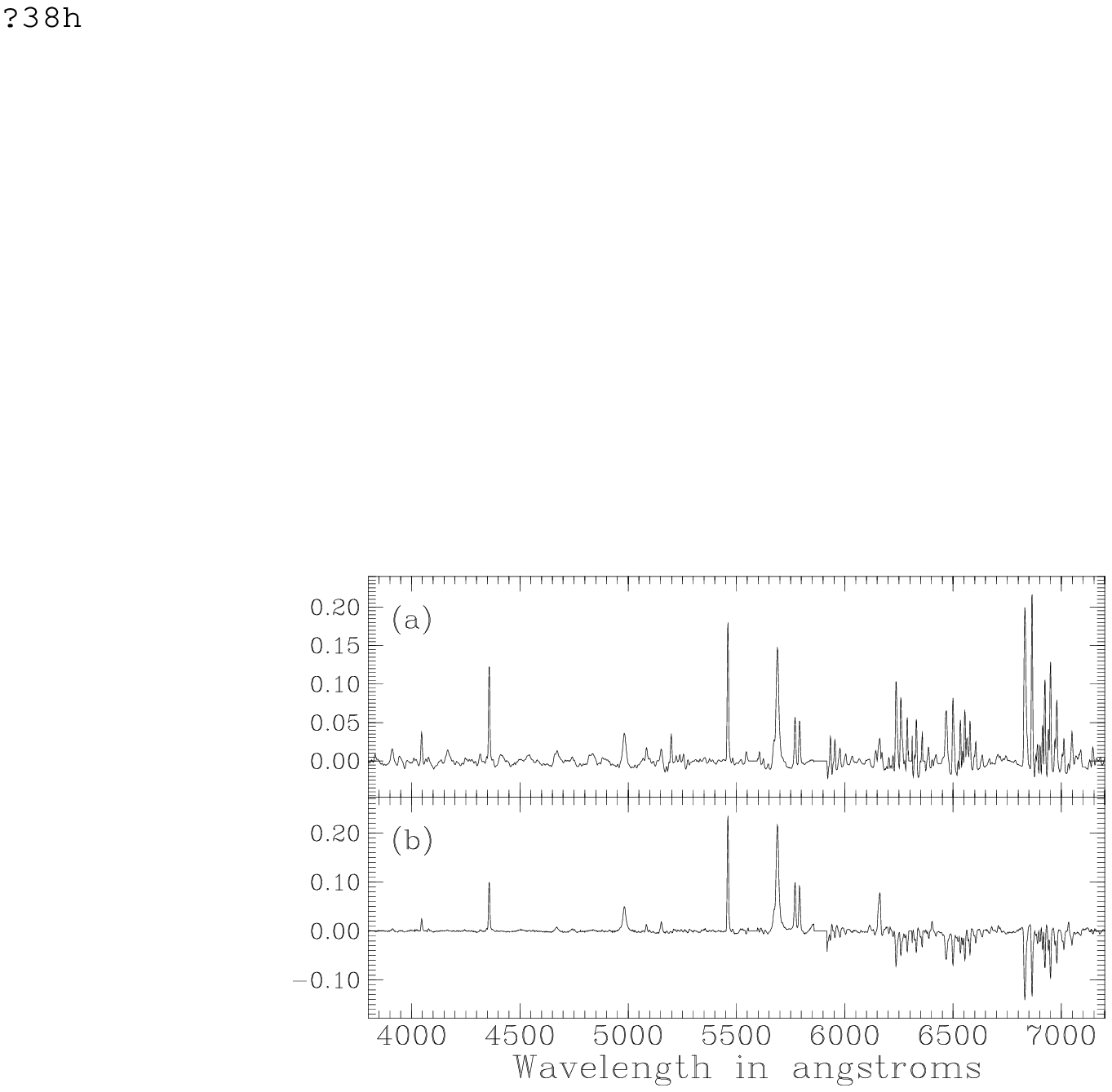}
\caption{\label{pass1vec}
The first two Eigenvectors in the continuum subtracted sky model.}
\end{figure}

Next we fit the residual sky $+$ object spectrum with a two component
Eigenvector model (section \ref{eigen}) of the continuum subtracted sky, and
subtract this fit.  Figure \ref{pass1vec} shows the two component sky
model used in section \ref{test}; to first approximation the first vector
represents the mean sky, and the second adjusts for the fluctuations
in the OH lines.  Typically we remove more than 80\%\ of the flux in the
residual sky spectrum with this step.

The second order residual which results from these steps typically
shows large (wavelength) scale variations due to errors in the
continuum estimation.  With most of the sky lines removed a second
order continuum fit may be made, resulting in a reflattened
(\cite{1991opos.meet..133K}) residual sky spectrum.

Finally we fit a several component (in section \ref{test} we use ten)
Eigenvector model of the reflattened residual sky spectrum, and
subtract it.

The residual from this last operation is the continuum subtracted
object spectrum, along with any remaining noise and errors.

\section {\label {eigen} The Eigenvector Model}

Eigenvector decomposition techniques (e.g. \cite{1924mmp1.book.....C},
\cite{1986nras.book.....P}) have long been used to classify
astronomical objects.  \cite{1964MNRAS.127..493D} used Principal
Component Analysis (PCA) to classify stars on the basis of multicolor
photometry; \cite {1973A&A....23..259B} used PCA to classify galaxies
on the basis of their integrated properties; and \cite
{1982PhDT.........2K} used PCA to classify stars on the basis of their
spectra.  In these cases classification is exactly equivalent to
fitting an Eigenvector model to the data, with the addition that the
resulting coefficient space is partitioned into classes.

More recently several authors have used PCA and Singular Value
Decomposition (SVD, a similar Eigenvector technique, see the
discussion in \cite{1986nras.book.....P}) to classify stellar
(\cite{1998MNRAS.298..361B} and references therein) and galaxy
(\cite{1998ApJ...505...25B} and references therein) spectra.

In this work we use SVD to create two Eigenvector models, first for
the continuum subtracted sky spectra, and second for the residual
reflattened sky spectra.  To accomplish this task we begin with a
set of well-observed sky spectra (we typically use 500 -- 1000), which
does not contain spectra with cosmic ray hits, bad focus, or other
anomalies.  We process these spectra so that each contains exactly the
same number of counts on exactly the same wavelength scale.
Additionally in all of these steps we ignore the pixels corresponding
to the sky lines we choose not to estimate, (section \ref{test}) OII
5007\AA, NaD, OII 6300\AA, and OII 6363\AA.

Next we subtract the continuum from each normalized sky spectrum using
a high order ($\sim$40) spline fit.  We decompose the continuum
subtracted, normalized sky spectra into Eigenvectors using SVD; the
first two Eigenvectors are our model for the continuum subtracted sky.
Note that higher order Eigenvectors are contaminated by errors in the
continuum fit.

Finally we fit the two component Eigenvector model for the continuum
subtracted sky to our set of normalized, continuum subtracted sky
spectra, and subtract the fit; then we reflatten the residual spectra
using another high order spline fit.  We then decompose the
reflattened residual spectra using SVD; the resulting Eigenvectors are
our final model for the reflattened residual sky spectra.

Because of the reflattening, the principal Eigenvectors in this final
model do not show contamination by errors in the continuum estimation;
thus there is not an obvious cut-off point for the number of Eigenvectors
to use in the fits.  This cut-off will vary from project to
project and from observing protocol to observing protocol.  In section
\ref{test} we use ten Eigenvectors; there is a clear change in slope
in the log-singular-value vs. Eigenvector-number diagram at the tenth
Eigenvector, indicating the transition from signal to noise.

\section {\label {test} Testing the Method}

The purpose of our new sky subtraction method is to allow observations
of fainter objects to be made with multi-object fiber-optic
spectrographs, such as the 300 fiber Hectospec (\cite
{1994SPIE.2198..251F}, \cite {1998SPIE.3355..285F}) on the converted
6.5m MMT.  A typical operating mode for this instrument is that each
half hour there would be three 10 minute exposures using the same
fiber set-up; the sets of three 10 minute spectra would be combined to
remove cosmic rays and other shot noise.  In this operating mode, with
10\%\ of the fibers devoted to sky observations, several hundred
defect-free sky spectra would be observed per night.

This instrument is not yet in operation; here we therefore build sets
of spectra for testing from the archive of spectra from the FAST
spectrograph (\cite {1998PASP..110...79F}).  FAST is a long slit
spectrograph residing on the 1.5m Tillinghast telescope.  We choose
two datasets, the first of well observed dark skys, both to create the
Eigenvector models and to add to the object spectra, and the second a
set of typical redshift survey spectra.

For the sky spectra we took all sky exposures (exactly 1000) made
between 21 Jan 1999 and 27 Dec 1999 where the exposure time was
greater than 500 seconds, the sun was at least 30$^\circ$ below the
horizon, the moon was at least 20$^\circ$ below the horizon, and the
instrumental set-up was close to that standardly used in redshift
surveys.  We removed spectra which showed signs of anomalies,
including cosmic rays or residual noise from UV flooding, for a final
list of 924 sky spectra.  The mean number of counts per 1.7\AA\ pixel in
these spectra is 569, with a factor of three variation about this
value.

We obtained these sky spectra from regions along the slit
adjacent to the objects being observed; they typically measured an
area on the sky of 3$\arcsec$ $\times$ 30$\arcsec$, which, for constant
observing time, produce a flux about three times what we expect
from a 1.5$\arcsec$ diameter fiber on the Hectospec/MMT, after taking
into account the 18.2 times larger collecting area of the MMT.  

The galaxy spectra are the set of all spectra from the 15R survey
(\cite {2000inprep....G}) where the exposure time was greater than 500
seconds, the sun was at least 30$^\circ$ below the horizon, the moon
was at least 20$^\circ$ below the horizon, and the redshifts were
judged secure by S. Tokarz following manual inspection. This dataset
consists of 1648 spectra, of $\sim$15th mag (R) galaxies, taken
between 1994 and 1997.  We then put these spectra through a fully
automatic redshift reduction (KM98), and remove the 15 spectra where
the fully automated result did not match the result confirmed by
manual inspection; typically this difference results from unremoved
cosmic ray hits in the spectra.  We further remove 110 spectra where
the $r$ statistic (a measure of quality of fit, \cite
{1979AJ.....84.1511T}) is $< 5$, the redshifts from absorption line
spectra with $r>5$ are nearly always correct(KM98); we routinely
accept them without manual inspection.  Our final test set is 1523
typical redshift survey spectra; each original spectrum is
reducible fully automatically.  The mean number of
counts per 1.7\AA\ pixel in these spectra is 277, the fluctuations
about this value are about a factor of 1.5.

We take each of the 1523 galaxy spectra, and combine it with ten
different sky spectra, which are created by summing between 1 and 10
randomly selected sky spectra.  This procedure yields a sample of
15230 test spectra with sky to object flux ratios which vary by more
than two orders of magnitude.

Each of these combined spectra then has the sky removed  by the
procedure described in section \ref{iter}, and has its redshift
determined by the automated methods of KM98.  We compare these redshifts 
with those derived from the original reduction of the galaxy
spectra; a redshift is considered correct if it is within 300$km/s$
 of the original value.

\begin{figure}[t]
\plotone{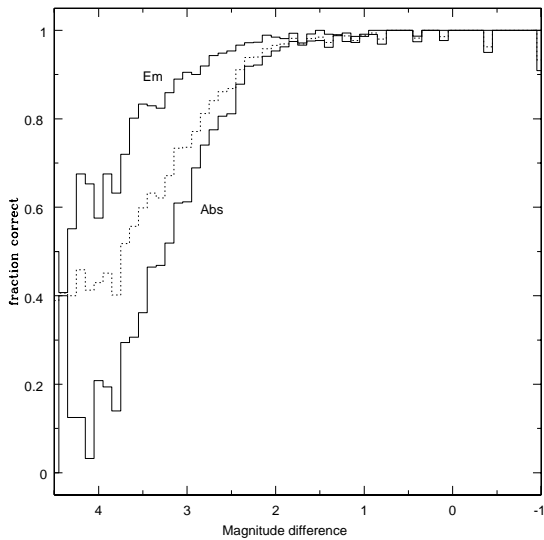}
\caption{\label{all3}
The fraction of correct redshifts 
as a function of the flux ratio of the object with the sky,
expressed in magnitudes.}
\end{figure}

Figure \ref{all3} shows the fraction of correct redshifts obtained in
this test as a function of the flux ratio between the object and the sky,
expressed in magnitudes.  This flux ratio refers to light entering the fiber,
and is thus primarily a measure of the seeing convolved central
surface brightness of the object with respect to the brightness of the
dark sky. 

We divide the galaxy spectra into two subsamples: emission line
dominated spectra, and absorption line dominated spectra.  The solid
lines marked em/abs represent spectra where the emission/absorption
line template (KM98) obtained the higher $r$ value in the original
reduction.  Emission line objects still give redshifts for objects
about a magnitude fainter than absorption line objects, everything
else being held constant.  The dotted line between the em/abs lines is
the total, and represents a typical mix of spectra for a magnitude
limited redshift survey at redshift 0.05 transformed to fainter
apparent magnitude.

The absorption line redshifts are more than 95\%\ correct until the
objects are 2 magnitudes fainter than the sky and more than 50\%\
correct until the objects are 3.25 magnitudes fainter than the sky.
The emission line spectra are more than 95\%\ correct until the
objects are 2.5 magnitudes fainter than the sky, and 50\%\ correct
until they are 4.4 magnitudes fainter than the dark sky.

To simulate the Poisson limit we add noise to the original spectra
corresponding to the Poisson noise expected from the number of counts
in the skies added to the spectra, using $\sqrt{n}$ for each pixel and
a Gaussian random deviate generator.  This procedure simulates perfect
sky subtraction.  We then put these spectra through the automated
processing of KM98 and obtain redshifts.

\begin{figure}[t]
\plotone{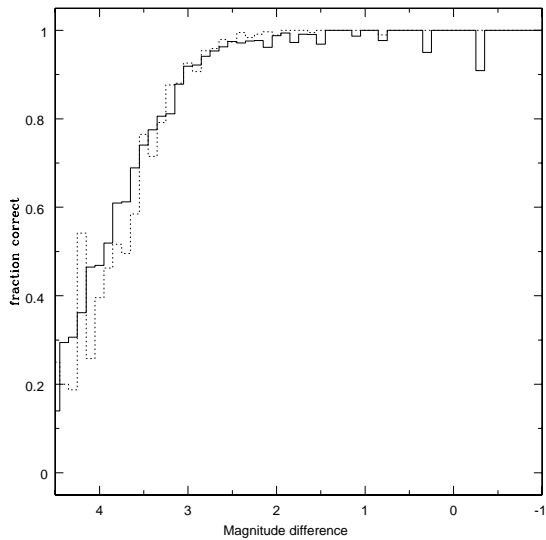}
\caption{\label{noiseab}
The fraction of correct redshifts obtained for Poisson noise limited
spectra, compared with the absorption line objets in figure \ref{all3}
shifted 0.7 magnitudes fainter}
\end{figure}

Figure \ref{noiseab} shows the fraction with correct redshifts as a
function of brightness ratio with the sky.  The dotted line shows the
fraction correct for the absorption line objects at the simulated
Poisson limit (i.e. with perfect sky subtraction); the solid line
shows the absorption line result from figure \ref{all3} shifted 0.7
mag fainter; the sky subtraction in this test came within 0.7 mag of
matching the limiting magnitude imposed by Poisson statistics.  The
plots for the emission line objects and the sum of all objects yield
very similar results.

\begin{figure}[t]
\plotone{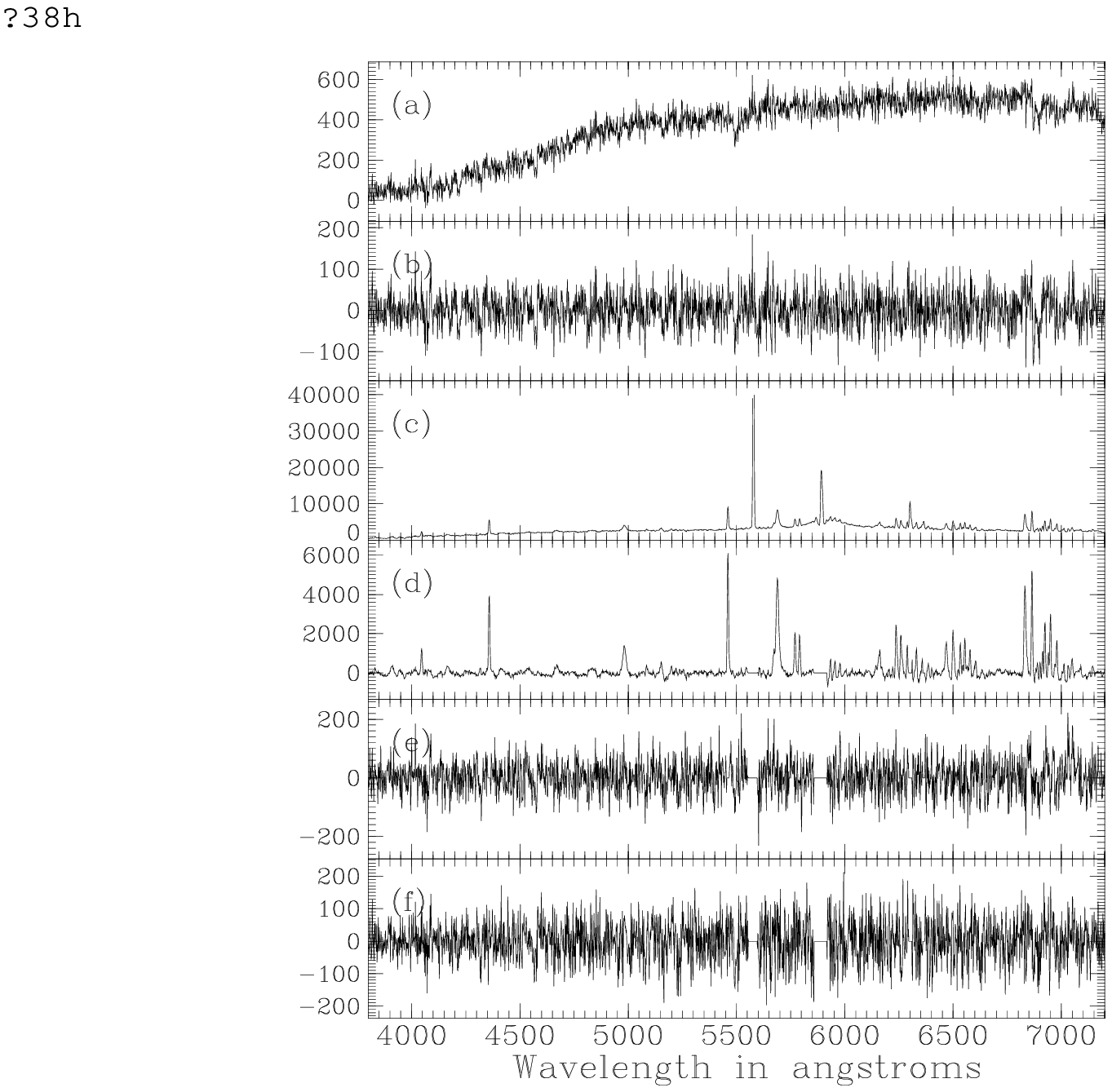}
\caption{\label{demofig}
A typical absorption line spectrum at various stages of the test; see text.}
\end{figure}

\begin{figure}[t]
\plotone{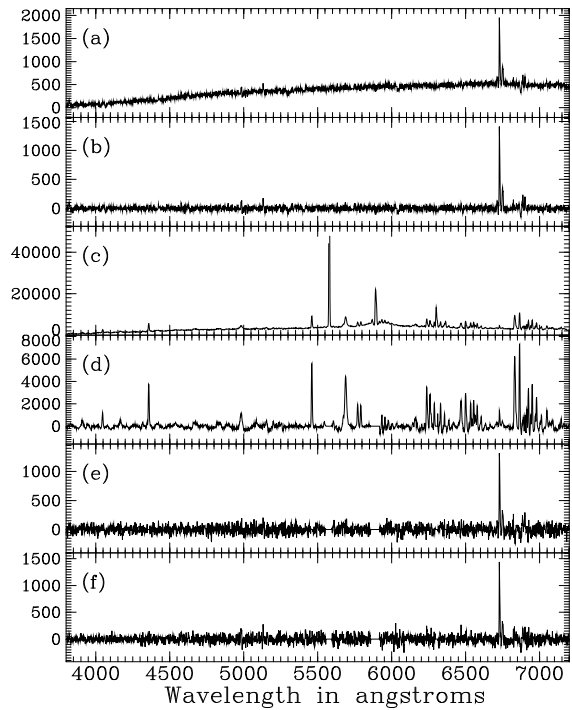}
\caption{\label{demo2}
A typical emission line spectrum at various stages of the test; see text.
THIS FIGURE WILL NOT BE PUBLISHED.}

\end{figure}

Figure \ref{demofig} shows a typical absorption line spectrum at
several stages in the test.  On top (\ref{demofig}a) is the original
galaxy spectrum from the 15R survey, its correlation with our
absorption line template gave an $r$ value of 7.6; below it
(\ref{demofig}b) is that spectrum with the continuum subtracted. Below
that (\ref{demofig}c) is the sum of that spectrum and a sky spectrum
which is the sum of seven original sky spectra, the object is 2.2
magnitudes fainter than the sky.  The next figure (\ref{demofig}d) is
the sky $+$ object with the bright sky lines and the continuum
removed.  Second from the bottom (\ref{demofig}e) is the residual
spectrum after the sky subtraction process, this spectrum yields the
correct redshift with an $r$ value of 4.2; the bottom plot
(\ref{demofig}f) shows the original spectrum in \ref{demofig}a with
added random Poisson noise corresponding to the number of sky photons
in \ref{demofig}c, and after the continuum has been removed.  This
spectrum yields the correct redshift with an $r$ value of 6.1.

THIS PARAGRAPH WILL NOT BE PUBLISHED. Figure \ref{demo2} is similar to
figure \ref{demofig}, but shows a typical emission line spectrum.  The
initial correlation with the emission line template gave an $r$ value
of 24.6; the object is 2.5 magnitudes fainter than the sky, and the
residual spectrum (\ref{demo2}e) gave the correct redshift with and
$r$ value of 18.3, while the Poisson limited noise spectrum
(\ref{demo2}f) gave the correct redshift with an $r$ value of 17.9.

\section {\label {discuss} Discussion}

Our new method for sky subtraction can remove the
spectrum of the sky from faint redshift survey spectra with an
efficiency within a factor of two of the Poisson limit, and {\it
without any simultaneous measurement of the sky spectra}.  

The sky spectra in the test were observed over one year, and contain
larger variations than would be found in a few-night Hectospec run.
For example the principal Eigenvector in the reflattened residual
model has NI 5199\AA\ as its strongest feature; this line was
substantially stronger during two months of the early summer than at
any other time.  Another Eigenvector, which shows a solar (zodiacal
light) spectrum, has a clear yearly variation.  Variations caused by
changes in the instrument set-up, focus, or chip response (we
periodically UV flood) will also affect the result.  Even with these
added sources of variance we obtain very good results for absorption
line spectra two magnitudes fainter than the sky, which if the
Mt. Hopkins sky is V$=$21.46 mag/arcsec (\cite{2000inpress.......M}),
and assuming a 1.5$\arcsec$ diameter fiber, corresponds to an object
with a 1.5$\arcsec$ aperture magnitude in V of 22.8; to measure this
would take about two hours of integration on the converted MMT.

Clearly the new techniques we have described can be combined with
simultaneous sky measurements to achieve improved results.  One
approach might be to fit a continuum subtracted mean spectrum derived
from the sky spectra in a single pointing (either derived from the
first Eigenvector, or some other method of obtaining a robust estimate
of the mean sky), then a fit to a set of reflattened residual
Eigenvectors, where the reflattened residual model is derived from a
much larger set of reflattened residual sky spectra, perhaps those
observed during an observing run.  This procedure would, in the first
pass, save for the continuum subtraction and the details of the
fitting, mimic current methods for simultaneous sky estimation.  The
second-pass fit with the reflattened residual model would allow
subtraction of correlated small amplitude sky flickering in the sky
lines, and systematic effects such as due to bending of the fibers.

With this combination of methods we expect that sky subtraction, save
for the unremovable Poisson noise component, will not be a dominant
limitation in deep observations with multi-object fiber optic
spectrographs.

We thank D. Fabricant, M. Geller, S. Tokarz, E. Falco, J. Roll, and
W. Wyatt for discussions.

Software implementing these techniques has been written in the IRAF
(\cite{1986SPIE..627..733T}) environment, and is available from the
authors.


\begin{thebibliography}{}

\bibitem[Bailer-Jones et al. 1998]{1998MNRAS.298..361B} Bailer-Jones, C. A. 
L., M.  Irwin and T.  von Hippel 1998. \mnras, 298, 361

\bibitem[Bromley et al. 1998]{1998ApJ...505...25B} Bromley, B. C.,
W. H.  Press, H.  Lin and R. P. Kirshner 1998.  \apj, 505, 25

\bibitem[Brosche 1973]{1973A&A....23..259B} Brosche, P. 1973.  \aap, 23, 259 

\bibitem[Courant and Hilbert 1924]{1924mmp1.book.....C} Courant,
R. and Hilbert, D. 1924, Methoden der Mathematischen Physik I,
Berlin:Springer Verlag

\bibitem[Deeming 1964]{1964MNRAS.127..493D} Deeming, T. J. 1964. \mnras, 127, 
493 

\bibitem[Fabricant et al. 1994]{1994SPIE.2198..251F} Fabricant, D. G., E. 
H. Hertz and A. H. Szentgyorgyi 1994. \procspie, 2198, 251

\bibitem[Fabricant et al. 1998a]{1998SPIE.3355..285F} Fabricant,
D. G., E.  N. Hertz, A. H. Szentgyorgyi, R. G. Fata, J. B. Roll and
J. M. Zajac 1998.  \procspie, 3355, 285

\bibitem[Fabricant et al. 1998b]{1998PASP..110...79F} Fabricant, D. , P.  
Cheimets, N.  Caldwell and J.  Geary 1998.  \pasp, 110, 79

\bibitem[Geller, et al. 2000]{2000inprep....G} Geller, M.J., et al
2000 in preparation

\bibitem[Humason et al. 1956]{1956AJ.....61...97H} Humason, M. L., N. U. 
Mayall and A. R. Sandage 1956.  \aj, 61, 97

\bibitem[Kurtz 1982]{1982PhDT.........2K} Kurtz, M. J. 1982. Automatic 
spectral classification. Ph.D. Thesis, Dartmouth Coll., Hanover, NH.

\bibitem[Kurtz and Mink 1998]{1998PASP..110..934K} Kurtz, M. J. and D. J. 
Mink 1998.  \pasp, 110, 
934

\bibitem[Kurtz and Lasala 1991] {1991opos.meet..133K} Kurtz, M. J. and
J.  Lasala 1991. in Objective Prism and Other Surveys,
ed. A.G.D. Phillip and A.R. Upgren, L.  Davis Press, 133

\bibitem[Massey and Foltz 2000]{2000inpress.......M} Massey, P. and
C.B. Foltz 2000,  \pasp, April 2000, to appear.

\bibitem[Press, et al. 1986]{1986nras.book.....P} Press, W. 
H., Flannery, B. P. Teukolsky, S. A. \& Vetterling, W.T. 1986,
Numerical Receipes, Cambridge: University Press

\bibitem[Shectman and Hiltner 1976]{1976PASP...88..960S} Shectman,
S. A.  and W. A. Hiltner 1976.  \pasp, 88, 960

\bibitem[Tody 1986]{1986SPIE..627..733T} Tody, D.  1986.  \procspie, 627, 733 

\bibitem[Tonry and Davis 1979]{1979AJ.....84.1511T} Tonry, J. and
M. Davis 1979.  \aj, 84, 1511

\bibitem[Watson et al. 1998]{1998fopa.proc...50W} Watson, F.,
A. R. Offer, I. J. Lewis, J. A. Bailey and K. Glazebrook 1998.  ASP
Conf. Ser. 37: Fiber Optics in Astronomy II, 50

\bibitem[Westphal et al. 1975]{1975ApJ...197L..95W} Westphal, J. A., J. 
Kristian and A. Sandage 1975.  \apjl, 197, L95-L98 



\end{thebibliography}
\end{document}